\documentstyle[prl,aps,multicol]{revtex}
\begin{document}

\begin{multicols}{2}

\noindent
{\bf Comment on "Observation of 2D Polarons and Magneto-Polarons on
 Superfluid Helium Films"}

A recent letter by Tress et al.~\cite{tress} found agreement between
electron mobilities on a helium film and the theoretical
mobility of a ripplonic polaron. We point out that this
agreement results from an incorrect determination of the holding field
and application of formulas for the isolated polaron to the case where
the polaron radius $R$ and scale of electron localization $L$
are limited by the Wigner-Seitz radius $r_{WS}$.  We show that the
experimental results are in disagreement with the theory of the binding
energy and dependence of polaron mobility $\mu_p$ on film thickness
and magnetic field.

The holding field is written as \mbox{$E_\perp^* = E_\perp + E_d$.} Here
$E_d$ is the sum of the fields of all primary image charges of
strength $\delta e$ in the substrate. $E_\perp$ includes the applied
field and the field of all polarization charges excluding the primary
images, \mbox{$\delta e + \delta_h e$}, beneath each electron. For a
saturated electron density $n$, $E_\perp$ is~\cite{kajita}
\begin{equation}
E_\perp = (1-\delta-\delta_h) ne/2\epsilon_0. \label{eperp}
\end{equation}
Here \mbox{$\delta = (\kappa-\kappa_h)/(\kappa+\kappa_h)$;}
\mbox{$\delta_h = (\kappa_h-1)/(\kappa_h+1)$;}
\mbox{$\kappa = 4$} and \mbox{$\kappa_h = 1.057$} are dielectric
constants of the substrate and helium, respectively.

The main cause of discrepancies with the results of Ref.~\cite{tress}
lies in this expression. An overestimate of $E_\perp$ leads to errors
in the binding energy and $\mu_p$. An underestimate of
\mbox{$L(0) \propto (E_\perp^*)^{-1}$} leads to an incorrect variation of
$L(B)$ [Eq.~(2) of Ref.~\cite{tress}] and hence mobility with
magnetic field.  We have calculated the image field by taking into
account the suppression of the film by the electron pressure and
finite separation of the electron from the surface of the film~\cite{hu}.

The characteristic binding energy is given by
\mbox{$B_0=(e E_\perp^*)^2/4\pi\sigma$.} Here $\sigma$ is the surface
tension.  Binding energies are smaller for the case
$ r_{WS} < k_c^{-1}$. Calculated values of $B_0$ for the densities
used are in the respective ranges 20~-~60~mK, 40~-~80~mK, and
190~-~240~mK for $n=$ 2.1, 3.0, and $6.6 \times 10^{13}$~m$^{-2}$
and are much less than $T=$ 1.3 K.  For $n=6.6 \times 10^{13}$~m$^{-2}$
the system is in the crystal phase.

The calculated polaron mobilities~\cite{es} are plotted versus the
suppressed film thickness $d$ and compared with experiment in
Fig. 1.  Vapor scattering is included by using Matthiessen's
rule as an approximation. The capillary constant $\kappa_c$ is
replaced~\cite{monarkha} by $2n^{1/2}$ in the expression for
$\mu_p$~\cite{tress} since \mbox{$k_c^{-1} > r_{WS}$.} We have not corrected
for \mbox{$L > r_{WS}$} which holds for some of the data. Theoretical curves
terminate at the limiting film thickness.

We compare the calculated conductivity ratio with experiment for
\mbox{$n = 5\times10^{13}/$m$^2$} in Fig. 2.  The transition from
$L(0)$ to the cyclotron radius is shifted to smaller values of $B$
because of the larger value of $L(0)$.  The expression for $L(B)$ does
not include electron scattering within the dimple which results in
diffusion of the cyclotron orbits and an increase in L.  Data for $d=40nm$
are within 2 percent of the melting temperature.

In summary, these results cannot be taken as conclusive evidence
of the polaron state.
\\

Supported by NSF grant DMR-REU-94-02647.
\\

\noindent
N.A. Rubin and A.J. Dahm

Department of Physics, Case Western Reserve

University, Cleveland, OH 44106-7079
\\

\noindent
PACS numbers: 67.40.Rp, 71.38.+I, 73.20.Dx, 73.50.Fq

\figure{Fig. 1.  Polaron mobility versus $d$. Curves are labeled:
         $n=6.6\times10^{13}$ m$^{-2}$ - $\Delta$ - solid line,
         $E_\perp=$ 228 kV/m;
         $n=3\times10^{13}$ m$^{-2}$ - $\bigcirc$ - dashed line,
         $E_\perp=$ 104 kv/m;
         $n=2.1\times10^{13}$ m$^{-2}$ - $\Box$ - dot-dashed line,
         $E_\perp=$ 72.6 kV/m.}

\figure{Fig.  2.  Conductivity $\sigma(0)/\sigma(B)$ vs field. Curves are
         theory: $d=$ 30 nm - $\Delta$ - solid line; $d=$ 40 nm
         - $\bigcirc$ - dashed line.}

\end{multicols}


\begin{thebibliography}{999}
\bibitem{tress}    O. Tress, Yu.P. Monarkha, F.C. Penning, H. Bluyssen,
                     and P. Wyder, Phys. Rev. Lett. {\bf 77}, 2511 (1996).
\bibitem{kajita}   K. Kajita, J. Phys. Soc. Jpn. {\bf 52}, 372 (1983).
\bibitem{hu}       X.L. Hu and A.J. Dahm, Phys. Rev. {\bf B 42},
                     2010 (1990).
\bibitem{es}       Eq.(\ref{eperp}) of Ref.~\cite{tress} should be divided
                     by $e^2$.
\bibitem{monarkha} Yu. P. Monarkha and V.B. Shikin, Fiz. Nizk. Temp.
                     {\bf 8}, 563 (1982) [Sov. J. Low Temp. Phys.
                     {\bf 8}, 279 (1982)].
\end{thebibliography}
\end{document}